\documentclass{article}
\usepackage{ijcai20}

\usepackage{float}
\usepackage{times}
\usepackage{soul}
\usepackage{url}
\usepackage[hidelinks]{hyperref}
\usepackage[utf8]{inputenc}
\usepackage[small]{caption}
\usepackage{graphicx}
\usepackage{amsmath}
\usepackage{amsthm}
\usepackage{booktabs}
\usepackage{algorithm}
\usepackage{algorithmic}
\title{A.I. based Embedded Speech to Text Using Deepspeech}

\author{
Muhammad Hafidh Firmansyah$^1$\and
Anand Paul$^1$\footnote{Contact Author}\and
Deblina Bhattacharya$^{2}$\And
Gul Malik Urfa$^1$\\
\affiliations
$^1$Kyungpook National University\\
$^2$École Polytechnique Fédérale de Lausanne\\
\emails
\{hafid, anand\}@knu.ac.kr,
deblina.bhattacharjee@epfl.ch,
theurfagul@gmail.com
}

\begin{document}

\maketitle

\begin{abstract}
	Deepspeech was very useful for development IoT devices that need voice recognition. One of the voice recognition systems is deepspeech from Mozilla. Deepspeech is an open-source voice recognition that was using a neural network to convert speech spectrogram into a text transcript. This paper shows the implementation process of speech recognition on a low-end computational device. Development of English-language speech recognition that has many datasets become a good point for starting. The model that used results from pre-trained model that provide by each version of deepspeech, without change of the model that already released, furthermore the benefit of using raspberry pi as a media end-to-end speech recognition device become a good thing, user can change and modify of the speech recognition, and also deepspeech can be standalone device without need continuously internet connection to process speech recognition, and even this paper show the power of Tensorflow Lite can make a significant difference on inference by deepspeech rather than using Tensorflow non-Lite.This paper shows the experiment using Deepspeech version 0.1.0, 0.1.1, and 0.6.0, and there is some improvement on Deepspeech version 0.6.0, faster while processing speech-to-text on old hardware raspberry pi 3 b+.
\end{abstract}

\section{Introduction}
\textbf{ASR}( Automatic Speech Recognition ) are directly mapped audio as input and the output as character using neural network\cite{BaskarEtAl:2019}.Nowadays, voice recognition service needs a user to pay for using the service, and users need pay with a range from four U.S. Dollar until one hundred and fifty U.S. Dollar for every four thousand requests. Google Assistant, Amazon Alexa are led on ASR system for various application \cite{DumpalaEtAl:2019}. Moreover, for another aspect need always connected with internet to process voice recognition, where \textbf{ASR} ( Automatic Speech Recognition ) provider using \textbf{API} (Application Program Interface ) for processing voice recognition, also user can’t modified what they need. Then, we choose deepspeech by Mozilla, which open-source voice recognition. That has excellent performance, also can use a big dataset using the English language for starting point, and also have many accents for the dataset, this paper based on \cite{DBLP:journals/corr/HannunCCCDEPSSCN14} Baidu research paper that publishes in 2016. This paper shows the performance deepspeech from the first version until latest version (0.6.0 version ) using raspberry pi three b+, as a device for implementation deepspeech which have low computational power will show advantages and disadvantages of each deepspeech version. This paper we will show the deepspeech sctructure on "\textbf{Deepspeech Neural Network Structure}" section, for details information deepseech model . advantages and disadvantages of each deepspeech version is show on "\textbf{Pre-trained data details}" section. Device details information that will be used on experiment will show on "\textbf{Device details}" section. For "\textbf{Experiment}" section provide data and result of the experiment on each deepspeech version. Final conclusion of the experiment section will explain on "\textbf{Conclusion}" section.

\section{Deepspeech Neural Network Structure}

Neural network are very important on deepspeech, neural network are work immitate human brain \cite{IslamEtAl:2019}. 

\begin{figure}[H]
\includegraphics[scale=0.5]{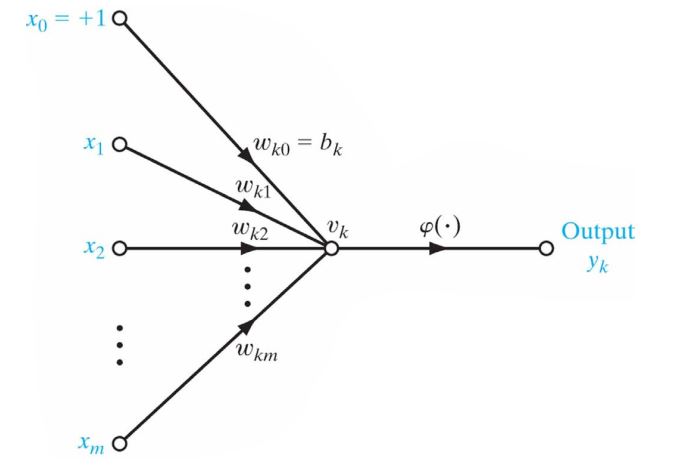}
\caption{Single Flow Graph Neuron}
\centering
\end{figure}
Figure 1 shows a single graph of the neuron on a picture, and formula its show on formula \textbf{(1)} section, where $\sum$ of $w_{kj},x_{j}$ and activation function that denotes by $\varphi$ will become a part of a neuron. From the simple neuron, it can be more complex neuron for processing input data.
\begin{align}
y_{k}=\varphi (\sum_{j=0}^{m}w_{kj}x_{j} )=\varphi (w_{k}^{t}x)
\end{align}
Deepspeech using \textbf{RNN} ( Recurrent Neural Network ), RNN will process process spectogram as input sequence into final transcript  as a readable text \cite{AmodeiEtAl:2015}.On neural network of deepspeech structure to ingest speech spectrogram for generating text transcript, have five hidden layers with three non-recurrent layers.Input denotes as $x$ , where hidden units for layer $l$ denotes with $h^{(l)}$ , $t$ as a time for sequence.The three layers using formula \textbf{(2)} to extract feature from the spectrogram.
\begin{align}
      h_t^{(l)}=g(w^{(l)}h_t^{(l-1)}+b^{(l)})
\end{align}
Using \textbf{ReLU} (Rectified Linear Unit ) for activation function, where ReLU $(x) = max(0,x)$ \cite{DittmerEtAl:2019}, $b^{(l)}$ and $w^{(l)}$ are the bias and weight for layer $l$ \cite{IakushkinEtAl:2018} , and layer four is a bidirectional recurrent layer, with have two parts, that is forward formula \textbf{(3)}.
\begin{align}
h_t^{(f)}=g(W^{(4)} h_t^{(3)}+W_r^{(f)} h_{t-1}^{(f)}+b^{(4)})
\end{align}
Backward formula on fourth layer denotes on formula \textbf{(4)}.
\begin{align}
h_t^{(b)}=g(W^{(4)} h_t^{(3)}+W_r^{(b)} h_{t+1}^{(b)}+b^{(4)})
\end{align}
for fifth layer will take forward and backward part as input using formula \textbf{(5)} and \textbf{(6)}: 
\begin{align}
h_t=g(w^{(5)} h_t^{(4)}+b^{(5)})
\end{align}
\begin{align} 
h_t^{(4)}=h_t^{(f)}+h_t^{(b)}
\end{align}
Output layer is standard softmax function for each time slice t and character k , using formula \textbf{(7)} and \textbf{(8)}
\begin{align} 
h_(t,k)^{(6)}=\widehat y_(t,k)={P}(c_t=k|x)
\end{align}
\begin{align} 
{P}(c_t=k|x)=\frac{\exp(W_k^
{(b)}h_t^{(5)}+b_k^{(6)})}{(\mathrm{\Sigma}_j\exp(W_j^{(6)}h_t^{(5)}+b_k^{(6)})}
\end{align}

\section{Pre-trained data details }

For pre-trained data, we didn’t training process again, but using same model that provide by deepspeech on each ver-sion, also using same one audio file for testing in each ver-sion deepspeech, the different only on environmtment where deepspeech are running like python version are matched by compability of each deepspeech version.

\begin{table*}[t]
\centering
\begin{tabular}{lrrrrrrr}
\toprule
Configuration & v0.1.0 & v0.1.1 & v0.2.0 & v0.3.0 & v0.4.0 & v0.5.1 & v0.6.0 \\
\midrule
train\_file     & - &LibriSpeech & LibriSpeech & LibriSpeech & LibriSpeech& LibriSpeech& LibriSpeech   \\
dev\_file 	    & - &LibriSpeech & LibriSpeech & LibriSpeech & LibriSpeech& LibriSpeech& LibriSpeech \\
test\_file      & - &LibriSpeech & LibriSpeech & LibriSpeech & LibriSpeech& LibriSpeech& LibriSpeech \\
train\_batch    & - & 12 & 24 & 24 & 24 & 24 & 128 \\
dev\_batch	    & - & 8 & 48 & 48 & 48 & 48 & 128\\
test\_batch     & - & 8 & 48 & 48 & 48 & 48 & 128 \\
epoch & -       & 13&30 & 30 & 30 & 75 &75 \\
learning\_rate  & - & 0.0001 & 0.0001 & 0.0001 & 0.0001 & 0.0001 & 0.0001  \\
dropout\_rate   & - & 0.2367 & 0.2 & 0.2 & 0.15 & 0.15 & 0.20 \\
n\_hidden       & - & 2048 & 2048  & 2048 & 2048 & 2048 & 2048 \\
\bottomrule
\end{tabular}
\caption{Comparison table on each version}
\label{tab:plain}
\end{table*}

\subsection{Deepspeech v0.1.0}

On deepspeech v0.1.0 didnt show the parameter that us-ing on  training process, then user must use their own con-figuration for training process, deepspeech v0.1.0 only pro-vide pre-trained data that contain lm.binary and trie file also alphabet list.  
Deepspeech v0.1.0 have some advantages , the advantages are : 
\begin{itemize}
\item The code is simple
\item Can run on raspberry pi
\item Provide model for testing
\end{itemize}
And also, there is some disadvantages on deepspeech v0.1.0 when running on production system, the disadvantages are :
\begin{itemize}
\item Didn't provide checkpoint to continue training
\item Prediction still not good
\end{itemize}
Deepspeech v0.1.0 accuracy also mention on \cite{ZengEtAl:2019} ,with the host transcript is  "\textit{I  wish  you  wouldn't}" but deepspeech v0.1.0 show the result "\textit{A sight for sore eyes}" , there is some miss on prediction text from an audio file.
\subsection{Deepspeech v0.1.1}
Deepspeech v0.1.1 show some information for training hyperparameter and also a changed that mention on the log information.There is only one change that happened on Deepspeech v0.1.1, initializing training run from the frozen graph. Deepspeech v0.1.1 provide model, alphabet list file and audio file.
Deepspeech v0.1.1 also has some advantages :
\begin{itemize}
\item Improvement on text prediction
\item Provide checkpoint for fine-tuning
\item Can run on raspberry pi
\item Deepspeech can run on python2.7 until the newest version
\end{itemize}
Besides the advantages of deepspeech v0.1.1, there are also some disadvantages on deepspeech v0.1.1, 
\begin{itemize}
\item Word Error Rate did not accurately
\item Using a huge amount of memory
\item Using a huge amount of CPU
\end{itemize}
Furthermore, for testing accuracy on deepspeech v0.1.1 already mention on \cite{ZengEtAl:2019} , the host transcript is  "\textit{I  wish you would not}". Furthermore, the result from the inference is "\textit{I wish you live}", there is some improvement, where deepspeech can predict the first two words from the host transcript.
\subsection{Deepspeech v0.2.0}
Deepspeech v0.2.0 didn't have much change , for the result on word error rate, based on Table 2 , still show achieved 11\% , but on deepspeech v0.2.0 itroduced new streaming API.
Deepspeech v0.2.0 advantages 
\begin{itemize}
\item checkpoint for fine-tuning
\item V2 Trie still use memory around 600 MB
\item Support for Node.JS
\end{itemize}
Deepspeech v0.2.0 disadvantages :
\begin{itemize}
\item Word error rate still high
\item Still use high amount of CPU and memory
\end{itemize}
\subsection{Deepspeech v0.3.0}
Deepspeech v0.3.0, based on Table 2, achieved word error rate 11\% on trained American English with LibriSpeech. Also, there is some change than the previous version. The most important on change are they fixed on memory leak and change on trie format.
the advantages of deepspeech v0.3.0 are :
\begin{itemize}
\item Provide checkpoint for fine-tuning 
\item Provide a model for testing
\item Support ARM64 device
\item Feature caching speed training
\end{itemize}
For disadvantages on deepspeech v0.3.0 are :
\begin{itemize}
\item V2 Trie still use a huge amount of memory , around 600 MB
\item Feature caching speed training still use more memory
\end{itemize}
Deepspeech also can implement on many language \cite{ArdilaEtAl:2019} , some languages can be achieved until 15\% for non-American English Language.
\subsection{Deepspeech v0.4.0}
Deepspeech v0.4.0 more stable than deepspeech v0.3.0 , it's achieved 8.26\% word error rate on Table 2 , word error rate on deepspeech v0.3.0 reduce by 2.74\%. Deepspeech v0.4.0 become version that will be ready for using TFLite ready model. 
The advantages of deepspech v0.4.0 are ,
\begin{itemize}
\item Feature caching speeds training
\item Have native client for Android
\end{itemize}
The disadvantages of deepspeech v0.4.0 are,
\begin{itemize}
\item Caching speeds training still use huge of memory
\item V2 Trie still use memory around 600 MB
\end{itemize}
\subsection{ Deepspeech v0.5.1}
Deepspeech v0.5.1 have more accurate for word error rate that only 8.22\%, and deepspeech v0.5.1 have testing demo for android device using tflite, and for the model is trained on 467356 steps.Deepspeech have compatible for running from checkpoint on deepspeech v0.5.0.
The advantages of deepspeech v0.5.1 : 
\begin{itemize}
\item Caching speeds training
\item Using TFLite 
\end{itemize}
The disadvantages of deepspeech v0.5.1 : 
\begin{itemize}
\item Still consume huge of memory for using caching speeds training
\item Code still not yet thread safe
\end{itemize}
\subsection{Deepspeech v0.6.0}
Dee pspeech v0.6.0 show some information for training hyperparamter and also some changed that mention on the log information.
	The improvement that happened on deepspeech v0.6.0 are :
\begin{itemize}
\item Change on data structure for language model trie file, can use memory mapped while loading.
\item Trie loading change become lazier , that reduce on memory usage and latency for first inference
\end{itemize}
And also on Deepspeech v0.6.0, they started using tensorflow lite for inference the process, it’s make more faster while doing the experiment on small raspberry pi 3 b+.
\begin{table}[H]
\centering
\begin{tabular}{lc}
\toprule
Deepspeech Version  & Word Eror Rate \\
\midrule
0.1.0   & - \\
0.1.1   & - \\ 
0.2.0   & 11\%\\
0.3.0   & 11\%\\
0.4.0	& 8.26\%\\
0.5.0   & 8.22\% \\
0.6.0   &\textbf{ 7.50\%} \\
\bottomrule
\end{tabular}
\caption{Word Error Rate on Each Version of Deepspeech}
\end{table}

On each version of Deepspeech, there is improvement quality on World Error Rate, it has shown by Table 2, for version 0.1.0 and 0.1.1 there is no precise information about the word error rate. For deepspeech version 0.3.0 the word error rate have same result with the previous version 11\%, but it's start going down again on next version of deepspeech, until it is reached version 0.6.0. From deepspeech version 0.2.0 the world error rate decreasing and become more accurate, it is indicated if the deepspeech have improved on accurate predicting text from  the audio file. 
\section{Device details}
On this section show what kind of device that using on the experiment, we use raspberry pi 3 b+ , we choose raspberry pi 3 b+ to implement pre-trained model data because raspberry can do computational processing without internet connection, and this is the detail of the device
\begin{table}[H]
\centering
\begin{tabular}{lrrr}
\toprule
Feature  & Model \\
\midrule
Model       & Pi 3 Model B+ \\
RAM         & 1 GB LPDDR2 SDRAM \\ 
CPU         &  Cortex-A53, SoC @ 1.4GHz\\
Ethernet    & Gigabit  Ethernet  over  USB  2.0  ( \\
WLAN	    & 2.4GHz,Bluetooth 4.2\\
HDMI        & Full-size \\
USB         & 4 USB-2.0 ports \\
Micro SD    & default \\
Power       & 5V/2.5A DC \\
\bottomrule
\end{tabular}
\caption{Feature and Model on Raspberry Pi}
\label{tab:plain}
\end{table}
The benefit for using Raspberry Pi is the device only use very small power consumtion that mention by \cite{BekarooEtAl:2016}, but we can explore for many things on the raspberry using full linux command, and also is possible to use all of GPIO pins for project.
\section{Experiments}
We do some experiment using deepspeech v0.6.0, v0.1.0 and v0.1.1. However, we do not do the test on deepspeech v0.2.0 until deepspeech v0.5.0 because there is some issue on compatibility environment on raspberry pi, which new modules become incompatible with some of deepspeech version. Deepspeech v0.2.0 until v0.5.0 must running on python3.4 and python3.5, but the module that we installed refused to run, then we did not continue the running test on deepspeech version 0.2.0 until 0.5.0.On each test, we use a model that already provide on each deepspeech version, and for the device that we use, we explain on the device section. On the experiment, we show the difference in memory usage, CPU and inference time, also how good each deepspeech version can recognize speech to text.
On this experiment, we use the waterfall method, which suitable with this experiment, and we already know about the process of this stage, on the waterfall it has six steps that need doing there are :
\begin{enumerate}
\item Requirements, on the requirements we collect for the initial phase, like make requirements document
\item Analysis, for this stage, we make model and logic for the experiment.
\item Design, this stage make technical design requirements, design the environment that suitable with each version of deepspeech and what environment that cant run on the deepspeech
\item Coding, on this step, we are building some script that can capture the result when the experiment on running,  
\item Testing, on this stage we test our system, to search a bug or check if the system on experiment already meet our requirements
\item The operation, the application that ready for doing on the experiment, on the operation we will get the result of each deep-speech
\end{enumerate}
\begin{table}[H]
\centering
\begin{tabular}{lrrr}
\toprule
Module & v0.1.0 & v0.1.1 & v0.6.0 \\
\midrule
Tensorflow 		& 1.14.0 & 1.14.0 & 1.14.0 \\
Numpy			& 1.15.4 & 1.16.5 & 1.17.4 \\ 
Scipy				& 0.19.1 & 1.2.2 &	- \\
H5py				& 2.10.0 & 2.10.0 & 2.10.0 \\
Deepspeech		& 0.1.0 & 0.1.1  &	0.6.0 \\
Python		& 2.7 & 2.7  &	3.7\\
\bottomrule
\end{tabular}
\caption{Comparison Module on Each Version}
\label{tab:plain}
\end{table}
Table 4 shows the difference and similarity of the module that installed on each version of deepspeech, and only deepspeech  version 0.1.0 and 0.1.1 that have same python environment
\subsection{Experiment Deepspeech v0.1.0}
Experiment for deepspeech v0.1.0 , using python2.7 as environmnet, because it’s only compatible with python2.7, will failed and show some errors if using python3 because the version build on python2.7 environment.

\begin{figure}[H]
\includegraphics[scale=0.5]{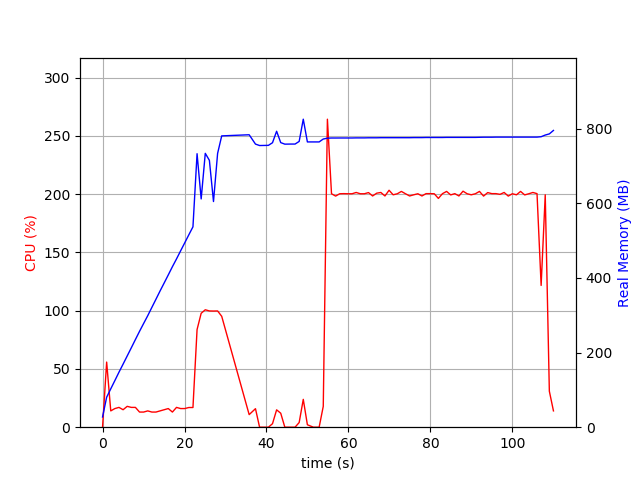}
\caption{Resource usage on deepspeech v0.1.0}
\centering
\end{figure}
On that Figure 1, show cpu resource usage will highly increase on 20 second and will reduce until 50 second, then will highly increase again until 250\% before 60 second and then stable on 200\% until inference process already done, on memorry usage, there is increasing from 0 to 800 MB , but in first process there are some spike on some seconds on the range 600 until 800 MB, inference time for one time process audio file is 102 second for 2 second audio file.
\begin{figure}[H]
\includegraphics[scale=0.85]{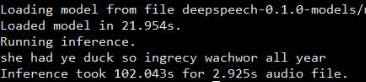}
\caption{Inference time on deepspeech v0.1.0}
\centering
\end{figure}
For the inference time, it's showing on Figure 2 ,loading time for model need up to 21 second until the result come out, and inference time need 102 second for 2 second audio file.
\subsection{Experiment Deepspeech v0.1.1}
On deepspeech v0.1.1 using python2,7 because its working well when running om python2.7 rather than on python3.

\begin{figure}[H]
\includegraphics[scale=0.5]{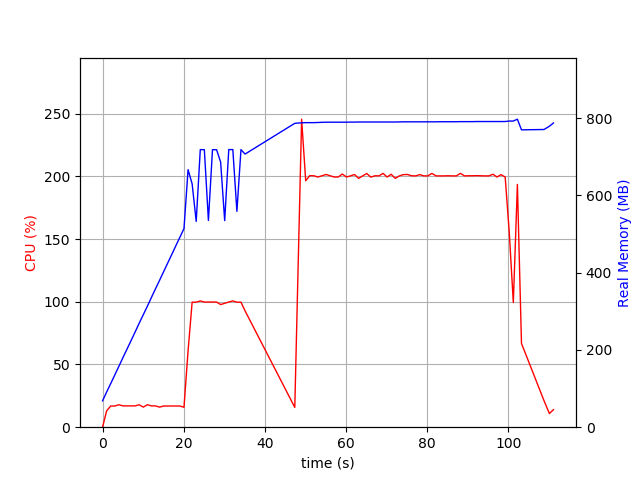}
\caption{Resource usage on deepspeech v0.1.1}
\centering
\end{figure}
On the Figure 3,the graph show cpu and memory usage on time scale (second), there are some increasing for memory usage from 200 MB until 800 MB , also there are some spike during 20 second until 40 second for memory usage , and then memory usage stable until the inference was done. For cpu usage there are some increasing usage until 100\% on 20 second and going down again on below 60 second, but cpu usage get increasing again until 250\% on 60 second and become stable on 200\% cpu usage until the inference was complete, on deepspeech v0.1.1 more better than deepspeech v0.1.0 , where deepspeech v0.1.1 more stable than preverious version, and there are some decreasing for time consuming.
\begin{figure}[H]
\includegraphics[scale=0.85]{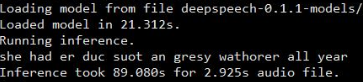}
\caption{Inference time on deepspeech v0.1.1}
\centering
\end{figure}
For the inference time, it's showing on Figure 4 ,for loading model need up to 21 second, and inference time need 89 second for 2 second sound, 13 second more faster for inference.
\subsection{Experiment Deepspeech v0.6.0}
On new version of deepspeech v0.6.0 , it’s need to use python3 ,because many dependencies of python3 more stable and better than python2.

\begin{figure}[H]
\includegraphics[scale=0.5]{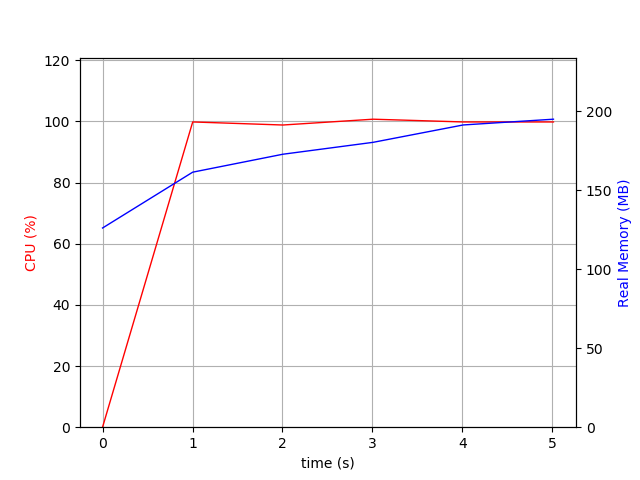}
\caption{Resource usage on deepspeech v0.6.0}
\centering
\end{figure}
Figure 5 show, the cpu and memory consuming on the graph, that graph show more stable on cpu and memory usage , on cpu usage from 0 second there is an increasing from 0 to 100\% and become stable until the inference completed, for memory usage there is an increasing from 100 MB to 200 MB until the inference already done, and only need 5 second to complete an inference.
\begin{figure}[H]
\includegraphics[scale=0.85]{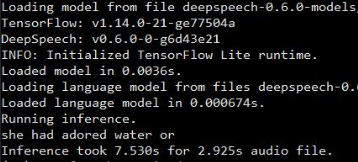}
\caption{Inference time on deepspeech v0.6.0}
\centering
\end{figure}
For the inference time it's showing on Figure 6 , for loading model only need 0,0036 second and for inference time need for 7 second.
\section{Conclusion}
Deepspeech v0.6.0 have very good implementation for embedded device like raspberry pi which didn’t have enough power to doing some complex computational, but the increasing performance of inference time from loading model until the result show deepspeech get increasing performance on embedded device like raspberry pi, the difference performance of each deepspeech version show on this table :
\begin{table}[H]
\centering
\begin{tabular}{lrrr}
\toprule
Resource & v0.1.0 & v0.1.1 & v0.6.0 \\
\midrule
CPU Usage ( \% )    & 250 & 250 & \textbf{100} \\
Memory Usage ( MB )	& 800 & 800 & \textbf{200} \\ 
Processing Time ( s )	& 102 & 100 &	\textbf{5} \\
Loading Model ( s )& 21  & 21  & \textbf{0,0036} \\
Inference Time ( s )	& 102 & 89  &	\textbf{7 }\\
\bottomrule
\end{tabular}
\caption{Comparison Resource Usage of Each Version}
\label{tab:plain}
\end{table}
On the table for maximum of cpu usage, for deepspeech v0.1.0 and deepspeech v0.1.1 still have same value on 250 \% , but for deepspeech v0.6.0 cpu usage going down until 150\% , and deepspeech v0.6.0 become more faster for processing. For maximum memory usage, deepspeech v0.1.0 and deepspeech v0.1.1 have same value on 800 MB , it’s need a huge memory consume for processing small audio file, and deepspeech v0.6.0 only consume 200 MB on peak time for processing same audio file on deepspeech v0.1.0 and deepspeech v0.1.1. Deepspeech v0.6.0 using less processing time, only need around 5 second, while deepspeech v0.1.0 need 102 second and deepspeech v0.1.1 need 100 second, this become deepspeech using less time and more faster in time sector. Time for loading model on deepspeech v0.6.0 is very fastly, only need 0,0036 second, while deepspeech v0.1.0 need 21 second and deepspeech v0.1.1 need 21 second.For inference time deepspeech v0.1.0 need 102 second to process , and deepspeech v0.1.1 more better which need 89 second, but deepspeech v0.6.0 only need 7 second to process same file with another deepspeech version. Based on the table 5 result and experiment that already done on raspeberry pi 3 , it’s show deepspeech v0.6.0 more faster and can be used on any device that have small computational power but enough to run computational algorithm, and no need internet connection for processing end-to-end speech to text.

\begin{table}[H]
\centering
\begin{tabular}{lrrr}
\toprule
Deepspech & Result  \\
\midrule
v0.1.0    & She had ye duck so ingrecy wachwor all year \\
v0.1.1	& She had er duc suot an gresy wahorer all year\\
v0.6.0	& She had adored water or \\
\bottomrule
\end{tabular}
\caption{Comparison Result of Deepspeech}
\label{tab:plain}
\end{table}
After processing using each model for deepspeech version, on table 6, it shows the different result on each version of deepspeech. The host transcript is "\textit{She had your dark suit in greasy wash water all year}", commonly all of deepspeech version successfully to process "she had" words, only deepspeech version v0.1.0 and version 0.1.1 have the result near of host transcript.
\appendix
\bibliographystyle{named}
\bibliography{bib}

\end{document}